# Somewhere Around That Number: An Interview Study of How Spreadsheet Users Manage Uncertainty

Judith Borghouts[a,b], Andrew D. Gordon[a,c], Advait Sarkar[a], Kenton P. O'Hara[a], Neil Toronto[a]

Corresponding Author: Judith Borghouts

[a]Microsoft Research, 21 Station Road, Cambridge CB1 2FB, United Kingdom
[b]UCL Interaction Centre, University College London, 66-72 Gower Street, London WC1E 6EA, United Kingdom
[c] School of Informatics, University of Edinburgh, 10 Crichton St, Edinburgh EH8 9AB

E-mail addresses: judith.borghouts.14@ucl.ac.uk (J. Borghouts), adg@microsoft.com (A. D. Gordon), advait@microsoft.com (A. Sarkar), keo@microsoft.com (K. P. O'Hara), netoront@microsoft.com (N. Toronto)

**ABSTRACT**
Spreadsheet users regularly deal with uncertainty in their data, for example due to errors and estimates. While an insight into data uncertainty can help in making better informed decisions, prior research suggests that people often use informal heuristics to reason with probabilities, which leads to incorrect conclusions. Moreover, people often ignore or simplify uncertainty. To understand how people currently encounter and deal with uncertainty in spreadsheets, we conducted an interview study with 11 spreadsheet users from a range of domains. We found that how people deal with uncertainty is influenced by the role the spreadsheet plays in people's work and the user's aims. Spreadsheets are used as a database, template, calculation tool, notepad and exploration tool. In doing so, participants' aims were to compute and compare different scenarios, understand something about the nature of the uncertainty in their situation, and translate the complexity of data uncertainty into simplified presentations to other people, usually decision-makers. Spreadsheets currently provide limited tools to support these aims, and participants had various workarounds.

## 1. INTRODUCTION

Data uncertainty is ubiquitous in various settings. Academics may deal with noise and missing data in their datasets, managers make business decisions based on projected sales data, and project leaders adapt schedules based on estimated workload. Having insight in uncertainty can help decide how reliable and how stable data is (Streit et al., 2008), can improve trust in the data (Joslyn and Leclerc, 2013; Sacha et al., 2016), and can help people make better informed decisions (Savage, 2010). Communicating uncertainty to the user however remains challenging, as it often requires the user to have an understanding of concepts such as probabilities (Kay et al., 2016; Savage, 2010; Tversky and Kahneman, 1974).

There have been several studies to identify people's strategies for dealing with uncertainty and reasoning about uncertainty (Boukhelifa et al., 2017; Schunn and Trafton, 2012), making a distinction between strategies that ignore, minimise, understand and exploit uncertainty. Furthermore, work has been done exploring different ways to represent uncertainty, such as data visualisations (Kay et al., 2016; Sacha et al., 2016; Streit et al., 2008) and textual statements (Joslyn and Leclerc, 2013). However, these bodies of work have two limitations: 1) strategies were described at a conceptual level, but not at a lower level to illustrate how specific tools are used to manage uncertainty; and 2) uncertainty presentations were either not evaluated with users (Streit et al., 2008), or they were evaluated with simplified tasks meant to study the perceptual and cognitive properties of the visual rendering. What remains unclear are the implications for people's actual work practices and how different material artefacts are used within these practices (Quinan et al., 2015).

For example, a widely used tool to manage and handle data are spreadsheets (Dourish, 2017; Grossman and Ozluk, 2009; Streit et al., 2008). Given the central role of spreadsheets in data practices, it is important to understand how uncertainty work is situated in spreadsheet practices within organisations. Understanding how the material features of spreadsheets are bound up in uncertainty practices can help identify opportunities where current tools might be reimagined to offer richer ways of supporting *uncertainty work*.

In this paper, we present an interview study with 11 spreadsheet users from a range of organisations and work domains. We examined the types of uncertainty and tasks they are involved in, what they use spreadsheets for, how they construct their spreadsheets and why they construct them in this way. Our contributions are: 1) a better understanding of the different roles spreadsheets play in individual and organisational use; 2) an insight into different strategies, difficulties and workarounds to manage uncertainty in spreadsheets; and 3) a discussion on how the design of spreadsheet tools could be improved to support managing uncertainty.

## 2 RELATED WORK

Prior work has looked at the type of strategies people use to cope with uncertainty (Boukhelifa et al., 2017; Schunn and Trafton, 2012), and has developed visualisations and models to express uncertainty (Joslyn and Leclerc, 2013; Kay et al., 2016; Streit et al., 2008). We build on this work and contribute added insight into how spreadsheets are used to manage uncertainty.

### 2.1 Strategies to cope with uncertainty

When people have to make decisions, they are generally trying to look for the most certain set of data to base their decisions on. Often however, there is a considerable level of uncertainty in data. Data may contain errors, missing data, or future estimates which are bound to change in the future. Tversky and Kahneman (1974) explain that people often find it cognitively difficult to comprehend uncertainty, and use several heuristics to simplify complex tasks that involve predicting uncertain values. This has also been found by more recent interview and field studies, which found that a common coping mechanism among scientists (Schunn and Trafton, 2012) and data workers (Boukhelifa et al., 2017) is to ignore, reduce, and minimise uncertainty, and that uncertainty may reduce people's trust in data (Sacha et al., 2016). However, in order to make informed decisions, it is important to have an understanding of the uncertainty of a situation. Savage (2010) uses the phrase 'flaw of averages' to refer to people's tendency to rely on a single average value rather than a range of possible outcomes, and warns that it can lead to a misconception of a situation and ill-informed decisions.

Boukhelifa et al. (2017) found that how workers cope with uncertainty is influenced by people’s goals, the types and sources of uncertainty, as well as their access to tools. They conducted an interview study with data workers, whose job revolves around gaining insight from large datasets, to better understand how they analyse uncertain data. Four high-level goals were identified: in addition to minimising and ignoring uncertainty, two other user goals were to understand and exploit uncertainty. While participants gave some examples of analysis tools they used, it was not explored in the study how these tools were used, and how the tools possibly influenced people’s coping strategies.

## 2.2 Uncertainty visualisations

To help people get insight into uncertainty, several studies have explored data visualisations. For example, Sarkar et al. (2015) used existing graphical representations of uncertainty as direct manipulation interfaces. By dragging on the ends of an error bar in a chart, users can specify how much uncertainty is acceptable for that element. This is particularly useful when the chart is an approximation, generated for instance by sampling from a larger database.

Uncertainty visualisations can be evaluated and analysed theoretically in terms of their perceptual and cognitive properties (Zuk and Carpendale, 2006). Zuk and Carpendale describe the application of different theoretical frameworks for evaluating uncertainty visualisations: the framework of Jacques Bertin (1983), which focuses on the structural use of graphic resources; the framework of Edward Tufte (1983), which emphasises Tufte's personal aesthetics and considerations as a craft practitioner; and the framework of Colin Ware (2004), which analyses the visualisation in terms of known psychological/cognitive and perceptual properties. Together, these frameworks can be used as a basis for heuristic evaluations (Nielsen, 1994) of uncertainty visualisations (Zuk and Carpendale, 2006).
A challenge in visualising uncertainty for everyday tasks however is a glanceability/false precision trade-off (Kay et al., 2016). On one hand, hiding uncertainty information from users can give a false sense of precision, but on the other hand, giving too much information can be overwhelming and confusing (Greis et al., 2017; Sacha et al., 2016). Some studies have evaluated visualisations with users to assess their comprehensibility using simplified tasks, but it is unclear whether these visualisations would be appropriate for more complex work practices. It is also worth noting that while visualisations of uncertainty have been given very sophisticated treatment in the context of statistical analysis, affordances for producing such visualisations in spreadsheets are scant. However, the liveness property (Tanimoto, 1990) of spreadsheets allows for rapid exploration of multiple values and scenarios. In this respect, the grid itself behaves as a simple 'visualisation' exhibiting desirable properties of Bertin's (use of the plane) and Tufte's (data-ink maximization, data density) frameworks.

### 2.3 Uncertainty in spreadsheets

Given the potential of data visualisations to capture data uncertainty, there have been proposals to place richer data types in spreadsheet cells, such as images, graphs, and visualisations of datasets. A common issue is that these implementations can quickly become too complicated for non-expert users to understand (Streit et al., 2008). Streit, Pham and Brown (2008) explored the advantages and possibilities of modelling uncertainty within spreadsheets, and developed a prototype that visualised uncertainty of spreadsheet data. While the prototype was briefly shown to domain experts, the emphasis of the study was very much on the technical possibilities rather than how it would be used by users.

Other tools have allowed users to run simulations and consider many different scenarios, such as Palisade @Risk and Oracle Crystal Ball. These tools allow the Monte Carlo methods (Glasserman, 2003) that are widely used in the finance industry to be directly performed within spreadsheets. Textbooks on spreadsheets, such as Winston (2011), exhaustively enumerate the features and techniques available, and describe how to perform Monte Carlo simulations without using add-ins. These tools are suitable for risk management, but are quite expensive to acquire and aimed towards users with a high level of expertise, both in the use of spreadsheets, as well as the formal theory of reasoning with probabilities. Consequently, they are not amenable to use by the vast majority of non-expert spreadsheet users.

### 2.4 The role of data tools in reflecting and shaping work practices

Though spreadsheets are commonly used for data analysis, very little work has been done on understanding how the properties and affordances of the tool are utilised by users to cope with uncertainty. When beginning to understand people’s actions, it is important to consider the tools that are used in the process, as the structure and forms of different genres of communication in organisations, be it a written report, presentation, or meeting, reflect and shape organisational practices (Yates and Orlikowski, 2007). In doing so, they can both enable and constrain action. To illustrate this argument, Yates and Orlikowski (2007) looked at how PowerPoint affects how people in organisations communicate. For example, PowerPoint presentations encourage textual content to be brief. Through empirical studies Yates and Orlikowski found that this was considered both enabling and constraining: it forced people to decide and focus on what was important, and slide decks were increasingly used over reports as deliverables to clients. However, abbreviated content could also lose its meaning if passed on to others unfamiliar with the activity. Though the typical form of presentations is a single person providing an oral explanation to a group of people, it was common to send the file to others, to be viewed in their own time.

Similarly, Dourish (2017) argues spreadsheets are situated in organisations in particular ways. The meaning in spreadsheets and their data emerges through the interactions that happen with and around it, in the context of the interactions that workers have with them. Uncertainty in spreadsheets is therefore not just a representational concern – any designs aimed at supporting uncertainty in spreadsheets should consider this broader context.

Spreadsheets both reflect work practices and shape them, as they allow for certain actions. Dourish highlights two key characteristics that define the structure and form of spreadsheets: the grid and formulas. Whatever content is being handled in a spreadsheet, the grid constrains the user to structure the data in a particular way, and forces people to decompose information into units which can be contained within individual cells. Through formulas, the user is able to define relationships between cells and both calculate and explore different values. In an empirical study looking at the role of spreadsheets in organisations, Dourish found that people specifically use spreadsheets because the grid structure matches a need for tabular data or lists, and formulas are used if data is expected to change in the future. The malleability of spreadsheets, that is the ease with which the content can be changed, makes them particularly useful in discussions when data is not final yet. Given the unique affordances of spreadsheets, we wonder how these affordances impact how people manage uncertainty.

While prior work has given insight into people's coping strategies, and different ways to visualise and compute uncertainty, it is unclear what strategies people currently use to cope with uncertainty in spreadsheets *specifically*. This makes it difficult to determine whether current spreadsheet tools adequately support users. The aim of our study was therefore to understand how people manage uncertainty in spreadsheets. Through interviews of participants where they shared and explained their own spreadsheets, we looked at people's use of spreadsheets to manage uncertainty in their everyday work practices. Concretely, our study aimed to answer the following research question: how are spreadsheets and its affordances used to deal with uncertainty?

## 3. STUDY DESIGN

### 3.1 Participants

We interviewed 11 participants (11 male) aged 26-72 (mean 38) from both industry and academia. They were recruited through a combination of convenience and snowball sampling. Participants worked in finance, construction, IT consulting, the oil and gas industry, business administration, and academic research. Participants were sent an email invitation and were eligible to participate if they used spreadsheets, which contained uncertain data. The invitation gave several examples of spreadsheet tasks which can involve uncertainty, such as budgeting, planning, business forecasting, data collection and analysis, scientific modelling, and making predictions. The invitation did not specify whether participants had to use spreadsheets for work or personal use, but all participants we recruited dealt with uncertainty in spreadsheets for work purposes. The size of participants' spreadsheets ranged from 40 rows to thousands of rows. Interviews lasted on average 60 minutes, and participants were reimbursed with a £30 voucher for an online store.

### 3.2 Procedure

The purpose of the semi-structured interview study was to better understand when spreadsheet users are faced with uncertainty and how they manage this within their spreadsheets. We asked participants to bring one or more example spreadsheets they used to the interview session. They were instructed to remove any sensitive information that they did not want to share. The interview took place at a location that was convenient for the participant, such as their home or office, our office, or a public café. Four interviews were conducted over Skype.

In the first part of the session, participants were asked to talk about their work, and how uncertainty and spreadsheets are a part of this work. We loosely based our discussion around five processes in handling uncertain data (acquire, manipulate, reason, characterise, present) taken from Boukhelifa et al's (2017) framework, to make sure we considered multiple processes during which participants may deal with uncertainty. In the second part of the session, we discussed if participants gain insights from uncertainty, what tools or strategies they use to gain this insight, and what challenges they perceive in doing so. In the final part, we asked participants to walk us through their example spreadsheets, explain how these spreadsheets were constructed, and what the spreadsheets were used for. The whole session was audio recorded, and participants' walk-through of their spreadsheets was screen recorded.

### 3.3 Data collection and analysis

The audio recordings were transcribed verbatim and analysed using iterative coding based on an inductive approach of thematic analysis (Braun and Clarke, 2006). There was no pre-existing coding scheme, but we did approach the data with a specific focus to uncover uncertainty types and insights, spreadsheet use, and strategies to manage uncertainty. Through a detailed inspection of the transcripts, we identified key features of the spreadsheets and work practices that related to participant concerns with uncertainty. In clustering related points in the interviews, an emergent thematic structure of high level concerns was derived and refined.

## 4 FINDINGS

In this section we start by describing the types of uncertainty and tasks participants dealt with. We then move on to how spreadsheet affordances were used and why, and the different roles that spreadsheets played in participants’ work and organisations. We end with a section linking identified uncertainty coping strategies with the type of uncertainty and type of spreadsheet role.

### 4.1 Types of uncertainty

Participants dealt with the following types of uncertainty: estimates, missing data, errors, dynamic data, untraceable and unfindable data. Estimates were the most common type of uncertainty among participants, and are approximated values of which the precise value is not known, such as expected profit: *'We're talking about the future. We don't know exactly what's going to happen. All we can do is make best estimates*' (P8). Missing data were values that were not recorded in the dataset, like gaps in measured sensor data, '*which are not necessarily explained*' (P11). In the case of errors, spreadsheet cells contained values that users believe to be incorrect based on their knowledge and expectations, or the cells contained an error message. Errors were caused by measurement errors, transcription errors, and broken links to external sources. Dynamic data refers to data that changes dependent on time, for instance if a spreadsheet shows stock market information of the current day or month: '*When you go to open that [spreadsheet] next month, the information has changed' (P9).* Untraceable data refers to data from which the source could not be traced: participants reported situations where it was unclear whether data they received from other people was derived from a computational model, or whether it was '*completely based on their intuition*' (P11). Lastly, unfindable data refers to situations where participants did not know how to find information in a spreadsheet. For example, P7 used timesheets which gave an overview of hours that all employees had worked per day. He wanted to know how many of these hours were worked on the weekend, but did not know how to retrieve this data from the timesheet, which made the weekend hours uncertain: '*There's a second unknown, which is the weekends. (…) Well for me it's difficult, I'm sure there's probably people that can extract it out there’* (P7).

**4.2 Types of tasks**

Participants dealt with uncertainty in spreadsheets for a range of tasks: they wanted to explore a space, get a better understanding of a situation, perform what-if scenarios, compare scenarios, make predictions and forecasts, perform sensitivity analysis, optimisation or constraint satisfaction, and wanted to get a better understanding of a situation for making decisions. We grouped similar types of tasks in three high-level purposes: analysing impact of scenarios, forecasting performance and comparing the impact of decisions. Table 1 specifies for each participant for which of these three purposes they used spreadsheets. Analysing impact of scenarios involved considering one or multiple possible things that could happen in the future, and the impact this would have on a particular entity, such as an organisation (P1, P3), the financial market (P3, P9), the climate (P2), a project (P4, P8) a building (P6), or a machine (P10, P11). An example of the impact of a scenario was given by P3: *‘If X wins the elections, it's likely that the market's going to react a little bit better, just because of her economic policies, therefore we would expect GBP to strengthen'* (P3). Forecasting performance involved making predictions on how a project, building or organisation would perform over time: *‘When we set the price for the project, we have to estimate how much it's going to cost us to build that project’* (P8). Lastly, participants used spreadsheets to compare the possible impact of different decisions. For example, P2 compared the impact of two different policies on tackling climate change: ‘*It doesn't necessarily tell you that either of them is the best thing to do, but it enables you to compare them (…) and then you can put in something else for policy A, (…) to see if that comes out better'* (P2).

| P# | Occupation | Impact of scenarios | Forecast performance | Impact of decisions |
|---|---|---|---|---|
| 1 | Independent accountant | | | • |
| 2 | Model developer | • | | • |
| 3 | Accountant for a bank | • | | • |
| 4 | Software engineer | • | | |
| 5 | Model developer | • | | • |
| 6 | Engineering consultant | | • | |
| 7 | Finance manager of public school organisation | | • | • |
| 8 | Financial risk manager oil and gas company | • | | |
| 9 | Spreadsheet designer for financial services | | • | |
| 10 | Research Associate | | • | |
| 11 | Research Fellow | | • | |

Table 1. The occupation of participants, and for which purpose each participant handled uncertain data in spreadsheets: they used them to analyse impact of scenarios, to forecast performance and to compare the impact of decisions.

### 4.3 People's choice of uncertain values

Participants constructed the spreadsheets themselves and knew which variables and values were uncertain, but did not always know 'how' uncertain they were: *'The final value will be somewhere around that figure. However, I haven't gotten to the point yet where I'm able to say: it's in this range, it's in the +/3% range'* (P8). All participants considered multiple possible values of these variables in their spreadsheets. As a spreadsheet cell only allows to input one value, several participants manually inputted and tried out different values in one cell, to see how this changed other values (P5, P6, P7, P8). This interaction with values enabled participants to understand the impact of a variable: *'If you change the glazing ratio five percent larger, then it had not much impact on the overheating of the building, but it had a huge impact on the embodied energy'* (P6, on analysing a building plan). Others saved these multiple possible values in the spreadsheet in different cells (P1,

P3, P4). They wanted to save these values to consider and compare multiple possible scenarios side by side. Common instances that were considered were the extreme minimum and maximum values, the mean value, or the most likely value. For most participants, their choice of values was influenced by their own expertise and expectations, the domain they worked in, discussions with colleagues and experts, the information they had on these variables and by looking at historical data. Participants with a higher mathematical expertise attempted to model uncertainty in their spreadsheets, and used their expertise knowledge to derive and calculate estimates from historical data (P2, P9, P10, P11): '*I would read as much around the literature as I possibly can, I will get in different views that people have (…) And then I'll work from all that to try and inform myself to be able to create a probability distribution that represents my view at that moment about the values for that parameter'* (P2).

**4.4 Comparison of scenarios**

Participants often used spreadsheets to compute and compare possible scenarios of a given situation. A spreadsheet was always an abstraction of some sort: it was infeasible to contain all possible scenarios, both due to limits of visual scaling as well as people's cognitive limitations to comprehend and compare many different scenarios. Therefore, users chose a subset to include which was based on their own judgment which scenarios were most likely or most important to consider, the domain they worked in, or based on specific instructions given by a boss or client. Participants inputted scenarios in spreadsheets by constructing the relationships between different variables. They then inputted different values for variables to see how this changed other values (P1, P5, P6, P7), or saved different scenarios in the spreadsheets and compared them side-by-side (P1, P2, P3, P4, P8). The specific properties of spreadsheets enabled this behaviour: the grid layout allows to put scenarios side by side, and formulas enable users to define and explore how cells are related. These spreadsheet affordances are discussed in more detail in the next section.

A common comparison was a best-case and worst-case scenario, but participants did not always consider a quantified likelihood of scenarios. They either did not have enough information available, or probability of scenarios was not well-understood and accepted in their domain. A qualitative strategy to validate likelihood of scenarios in these cases was to compare their scenarios with situations from the past (P3, P9), or discuss it with experts or colleagues who were not involved in the analysis process (P3, P6). Some participants did input uncertain values as distributions to look at the distribution of outcomes, and view the probability of scenarios occurring (P2, P10, P11).

### 4.5 Use of spreadsheet affordances

Participants worked with spreadsheets that contained a mixture of certain and uncertain data. Spreadsheets currently have no visual cue to differentiate between these different kinds of data: both data types are shown as deterministic numbers. Participants therefore used other spreadsheet properties to make a visual presentation of uncertainty within spreadsheets: they used the grid to spread out possible values over multiple cells, inputted formulas rather than discrete values, added written instructions to explain the source of uncertainty, coloured cells to distinguish different levels of uncertainty, and used different sheets to separate certain and uncertain data.

The grid layout of spreadsheets was used to juxtapose scenarios next to each other. One scenario was either laid out horizontally across one or multiple rows, or vertically across one or multiple columns. Participants placed scenarios next to each other to be able to visually compare them. If it took up too much space in a spreadsheet to easily compare all parameters of scenarios, participants worked out one scenario per worksheet. They then however still made a table to compare the outcomes of scenarios side-by-side, as shown in Figure 1. In this figure, the participant has made a table to compare the total costs of a decision for different financial years (e.g. FY18, F19) in a best-case and a worst-case scenario.

Comparison of scenarios also happened outside spreadsheets. Three participants manually changed values in the spreadsheet that were uncertain, to see how changing these values influenced other values. They memorised this change in values and made a comparison in their mind, or they wrote down the results on a piece of paper (P1, P4, P10).

| | A | B | C | D | E | F | G | H |
|---|---|---|---|---|---|---|---|---|
| 1 | | | | | | | | |
| 2 | | **Total funding cost** | | **FY18** | **FY19** | **FY20** | **FY21** | **FY22** |
| 3 | | **Best case** | | 38,935,872 | 35,646,389 | 33,147,656 | 32,342,537 | 30,801,791 |
| 4 | | **Worst case** | | 50,079,478 | 45,329,981 | 41,488,056 | 40,037,894 | 37,840,434 |
| 5 | | | | | | | | |

Figure 1. Participants used the grid layout to juxtapose and compare different scenarios. In the example shown here, the participant wants to compare the costs of a decision over different financial years in a best-case and worst-case scenario.

Formulas were used to interact and play around with different values, and to see how changing the value of one variable influenced the values of others. It was also used to indicate in the spreadsheet that a value was uncertain, and was dependent on the values of other cells. Figure 2 shows an example of a spreadsheet where the participant tries to determine the total number of days a project will take to complete. He has broken down the project in subtasks, and has estimated the number of days for each subtask: in the figure, these estimates are shown in cells B6 to B10. By adding these days together, he estimates that the project will take nine days, as shown in cell B11. However, because these individual estimates are uncertain and can change, rather than inputting the number 9, he uses a SUM formula to add up the estimates, so that the estimate for the project will automatically be updated if one of the other estimates change. Another common use of formulas was to look up certain values from a dataset. Three participants defined uncertainty in spreadsheets as formulas from which it was uncertain whether it retrieved the data they were looking for: '*It's all a bit like, I have no idea if this is doing anything right. (…) we ended up almost running a foul on the project where we realised it [formula] only looks data up if it's organised in a particular way*' (P6) Comments were only used if spreadsheets were shared with others. Participants who used models to make forecasts created an instruction sheet with a written explanation on how numeric values on the other worksheets were derived (P5, P6).

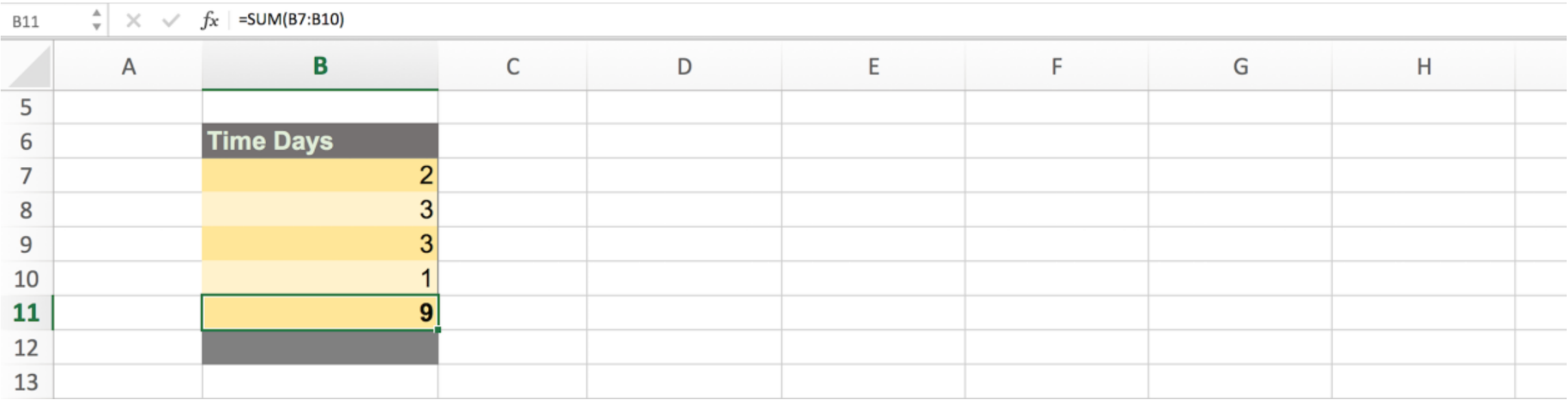

Figure 2. Participants used formulas, rather than discrete values, to indicate a value was uncertain and dependent on the values of other cells. In this example, the participant sums up the number of estimated days to complete individual tasks (cells B6 to B10), to determine how many days a project will likely take to complete in total (cell B11).

Five participants used colour in their spreadsheets to indicate the level of uncertainty (P1, P2, P4), and whether values were uncertain or not (P5, P6). Though there was a trade-off with representational accuracy of uncertainty by representing it through colours in comparison with exact numbers, there was a visual immediacy to it which was quicker to cognitively process than numbers. By using colour, participants were able to have a visual overview of how uncertain their dataset was, and could quickly identify which values were more uncertain than others: *'Instead of having to scroll through each cell and looking at the actual value, you can tell it to conditionally format to green is low, red is high, and it'll pick up quite easily where things might look a bit funny, which is useful'* (P6). Colour was also considered a useful tool to communicate uncertainty to others, who had not been involved in the data analysis process and were less familiar with the data. For example, P1 used green, yellow and red to communicate to clients how reliable the values were: *'What we've done is try to show them [clients] the levels of trust they can take in a particular category of data. (…) There's a visceral immediacy to it, it's traffic lights'* (P1).
Figure 3 shows an example of use of colours. The participant has estimated the number of days tasks will take him to complete. For each estimate, he has added a percentage of how confident he is about this estimate, and has coloured this cell either green if it is above a certain percentage, and red if it is below a certain percentage.

| | A | B | C | D | E | F | G | H |
|---|---|---|---|---|---|---|---|---|
| 14 | | | | | | | | |
| 15 | | Element | Code | Nature of Work | Time Days | Dependencies | Confidence % | |
| 16 | | Main Site | MS::C | Content | 2 | None | 80 | |
| 17 | | Vocabulary Builder | C::VB::C | Content | 3 | None | 10 | |
| 18 | | | | | | | | |

Figure 3. Participants coloured cells to indicate the level of uncertainty. In this example, the user has expressed his confidence regarding estimates of number of days to complete tasks.

Three participants made deliberate use of different worksheets to distinguish between certainty of data, and placed certain and uncertain data on different worksheets (P2, P7, P8). They did try to keep these sheets in the same workbook, because they often used certain data to calculate uncertain data values through dependent formulas.

**4.6 Need for certainty in decision-making**

Data went through several stages and transformations within the organisational workflow, and participants encountered uncertainty during data collection, analysis, and presentation. How data was manifested in spreadsheets depended on what stage in the workflow and decision-making process the spreadsheet was used. The closer people were to the decision-making process, the more need there was for certainty. There was also social pressure and expectations within certain domain cultures to present certainty: '*The more senior you are, the more you want an answer, either yes or no, there's not really a thing in-between*' (P3, accountant at bank). An important task for an analyst was to collect, analyse and transform uncertain data into a simplified presentation that could be used by the decision-maker. However, knowing how to simplify it, and what aspects to present and exclude, required a sufficient understanding of the uncertainty.

The first stage was data collection. When participants acquired past data that contained uncertainty, such as errors or missing data, this data had to be transformed and prepared for analysis. Common strategies were to understand the reason of uncertainty by evaluating the data source and acquiring more data, or to reduce uncertainty by removing and replacing these values.

The second stage was data analysis, the stage where participants aimed to understand the level of uncertainty of their data. They acquired this understanding by comparing multiple scenarios, inputting multiple possible values, discussions with colleagues, and comparing the data with historical data as well as their own expectations. These strategies were used as tools to determine what key values best captured the data, and were suitable to present.

In the final stage, data was transformed in a simplified version to present to decision-makers, such as senior management in the organisation, or clients. Uncertain data had to be transformed into certain, fixed values to make eventual decisions: '*You've always got to run it under uncertainty. (…) But then (…) the best representation of this whole very uncertain world we're in, is the mean numbers. If we're going to design policy, we should probably design it around those*' (P2)

Even though decision-makers required simplified data, there was an implicit understanding that this data was not incomplete or that data had been thrown away, but rather that it had been transformed, and the final simplified representation was underpinned by insights from data collection and data analysis: *'You will do all of the background analysis, (..) and you walk in there with a simple decision for them. Come in there with eight, and they'll look at you and go, 'Well I'm not doing that, that's your job to filter that'* (P1).

### 4.7 Trust, transparency, and credibility

An important aspect in presentation of uncertainty was to convey trust with the audience, which was considered more important than accuracy: '*I was quite precise, but it didn't agree with his expectations, he didn't trust what was coming across*' (P1). Depending on the domain, audience and their understanding of uncertainty, presenting uncertainty could increase or decrease trust*: 'The only way to be honest about it is to put the uncertainty in there'* (P2, presenting climate change predictions to policy makers). *'It [uncertainty] shows ambiguity, and people just don't like that'* (P3, presenting financial forecasts to bank). To convey trust, participants needed a narrative to support results and a spreadsheet played an important role in constructing a narrative. It improved the user's understanding of data and thus their confidence when presenting results to others. Spreadsheets also served as evidence to present if audience did not trust results. Participants improved trust in estimates by discussing and validating these with experts and colleagues. Another factor of trust was the credibility of the presenter: '*I've had roles previously where the credibility's built up enough so I can give someone a number and they won't question it, and it'll be fine*' (P3).

### 4.8 Presentation of (un)certainty

Apart from P4 and P11, participants were not involved in decision-making themselves, but were primarily responsible for analysing data and presenting results to clients and bosses who would make decisions. A challenge for participants was to translate the uncertainty from complex spreadsheets into simplified values for their audience to understand it. There was a need for certainty to make decisions, but participants highlighted that it was important to still understand uncertainty surrounding these numbers to make an informed decision. Often, there was little appreciation for uncertainty from their audience, and participants were expected to present fixed numbers. Strategies to communicate uncertainty to others were to visualise data through graphs, presenting users with interactive models, or give a qualitative explanation that numbers should be seen as uncertain: *'I do provide an estimate which is a hard figure. But what I tell [clients] is: it's around that figure'* (P8). P2 and P5 presented interactive models, so users could change values themselves and understand the uncertainty: *'That's the advantage of having a model, rather than a report. You can (…) give it to them and they put their own new things in there'* (P2).

### 4.9 Spreadsheet roles

Spreadsheets were used as (1) a database, (2) a template, (3) a calculator and analysis tool, (4) a notepad, and as (5) a data exploration tool.

A first type of spreadsheet use was as a database to store historical data. In this role, it was important to find specific data and spreadsheets were used within the organisation as a shared record*: 'The vast bulk of users use it as a way of storing, analysing, capturing data. (…) it becomes the record of the corporate memory'* (P1). Spreadsheet content was used and re-used to build scenarios and make more accurate estimates of uncertain values. An issue with using spreadsheets as a database was that datasets were often too large to be contained within one spreadsheet, and had to be split up (P3, P9, P10, P11). Data in spreadsheets was also not easily searchable. Participants used formulas to look up certain values in a grid, but participants mentioned they did not know how to check that these formulas were retrieving the correct data (P6, P7). This introduced additional uncertainty in their spreadsheets.

Spreadsheets also served as a record of past activity, and a second type of spreadsheet use was as a template to repeat analysis. While spreadsheets recorded the analysis process, the structured data and captured the relationship between different variables, they did not contain the actual outcomes and decisions that were made based on this analysis, which had to be recalled from other sources: *'Even sometimes when you use past examples, you still think: There's still something not quite right about that figure'* (P4).

A third type of spreadsheet use was as a calculation and analysis tool. In this role, it was important to work with accurate estimates of uncertainty. P2 and P5 used spreadsheets to build prediction models. Rather than using fixed values, uncertain data was often inputted as a formula, which was derived from other values in the spreadsheet. Use of formulas was important to ensure users were working with the most up-to-date values. Participants also inputted different values, to see how changing these values changed the other values in the spreadsheet (P4, P5, P7).

A fourth type of spreadsheet use was as a notepad, for example to type and work out different scenarios. In this role, it was important for users to visually see which cells were uncertain. Spreadsheets were used for an individual user to get a better understanding of the situation and reflected one user's point of view. Participants used this knowledge in discussion with colleagues and bosses, and in negotiations with customers and clients. For example, P4 used spreadsheets to get a better understanding of how long a project would approximately take. He would make different estimates and see how this influenced the total costs and/or expected completion time. Based on this, he made an estimate of costs and used this knowledge to take into a negotiation with the client, but did not take the actual spreadsheet with him.

Lastly, spreadsheets were used as an exploration tool to get a better understanding of a dataset and the uncertainty of a situation. P10 and P11 dealt with large datasets and used statistical analysis tools to analyse these. However, rather than importing these datasets into the analysis tools directly, they would first import these datasets into spreadsheets to get a visual overview of the dataset in a grid structure. In this way, spreadsheets facilitated the sensemaking process (Russell et al., 1993), as they enabled users to instantiate many potential representations of their data, and therefore to search for an appropriate representation of their data.

### 4.10 Collaboration for spreadsheets and spreadsheets for collaboration

Participants usually worked on the construction and editing of spreadsheets alone, and were the sole author and owner of a spreadsheet. A common reason was that the structure of spreadsheets could get complex with a lot of dependent and nested formulas, and it was difficult to communicate the structure to others: *'That [spreadsheet] is quite difficult for somebody else to disentangle. (…) The easiest way is almost to make a new spreadsheet doing it your own way'* (P6). There were also concerns that the spreadsheet was breakable if the structure of the spreadsheet was not well-understood*: 'I wouldn't trust anyone else to not break [the spreadsheet] [laughs]. But I will normally share the result'* (P11). If a spreadsheet was shared, complex formulas would be hidden or simplified to make it easy to understand (P5,

P7), and the construction and complexity was communicated through colour, and written comments and instructions (P2, P5, P10).

If spreadsheets were used as a notepad, a spreadsheet represented the point of view of a single user, and content was influenced by datasets and their own judgment. If spreadsheets were used as an analysis tool, collaboration on content was more common. Participants worked with colleagues and experts to decide on estimates, which scenarios to analyse, and whether assumptions made for making estimates were correct. Depending on the experience and credibility of the user, this discussion was sometimes also necessary to validate and justify values. Despite collaboration on the content, there was still one author who had ownership of the spreadsheet, and other collaborators did not have access to edit spreadsheets directly. Collaboration and communication happened outside of spreadsheets, through collocated discussions, emails and phone calls.

### 4.11 Linking uncertainty strategies with types of uncertainty and spreadsheet role

The above sections described the types of uncertainty, people's strategies to manage uncertainty and the roles spreadsheets played in people's work. In this section we try to link these together, to explore the relationship between spreadsheet use and people's strategy goals, as well as the relationship between spreadsheet use and types of uncertainty.

We used Boukhelifa et al's (2017) framework as a starting point to cluster and label all of the strategies we found from our analysis. Each strategy was categorised according to type of uncertainty, and labelled to specify the high-level goal of the strategy: minimising, ignoring, understanding or exploiting the uncertainty. In the original framework, uncertainty coping strategies were then categorised according to different processes of data analysis. Because we were interested in people's strategies in spreadsheets specifically, we instead categorised the strategies according to different spreadsheet roles.

We found that the role a spreadsheet played in people's work influenced the types of uncertainty in that spreadsheet, and that people had different strategies for different spreadsheet roles. These findings are illustrated in Figures 4 and 5.

The bar chart shown in Figure 4 illustrates the number of times a strategy was categorised according to type of uncertainty, clustered per spreadsheet role. The figure highlights that, overall, participants primarily dealt with estimates in their spreadsheets, especially when using spreadsheets as an exploration tool and notepad. The only spreadsheet use in which participants did not deal with estimates was when a spreadsheet was used as a database. In this role, spreadsheets contained uncertainty in the form of unfindable data, missing data, errors and dynamic data. Overall, the figure illustrates that people dealt with different types of uncertainty, depending on the spreadsheet role.

The bar chart in Figure 5 shows the number of times a strategy was categorised under different strategy goals, clustered per spreadsheet role. What this figure shows is that people primarily tried to understand and minimise uncertainty. When used as a notepad, participants mostly tried to understand uncertainty, for example by reading additional literature, and discussing it with colleagues. At this stage in their work, spreadsheets were primarily used to write down and compare different scenarios. However, when used as an exploration tool, participants both tried to understand but also minimise uncertainty, for example by acquiring more data or replacing estimates with deterministic values. At this stage in their work, participants were usually dealing with a specific dataset.

The figure further shows that participants only tried to exploit uncertainty if spreadsheets were used as a calculation tool, exploration tool or notepad. For example, they exploited uncertainty by interacting with a model and playing around with many different values. Overall, the figure shows that people had different goals of how to deal with uncertainty, depending on the spreadsheet role.

Table 2 shows the full overview of all uncertainty coping strategies, categorised according to spreadsheet role and type of uncertainty, and labelled according to high-level goal. Any new strategies we found compared to the original framework of Boukhelifa et al. are highlighted in italics.

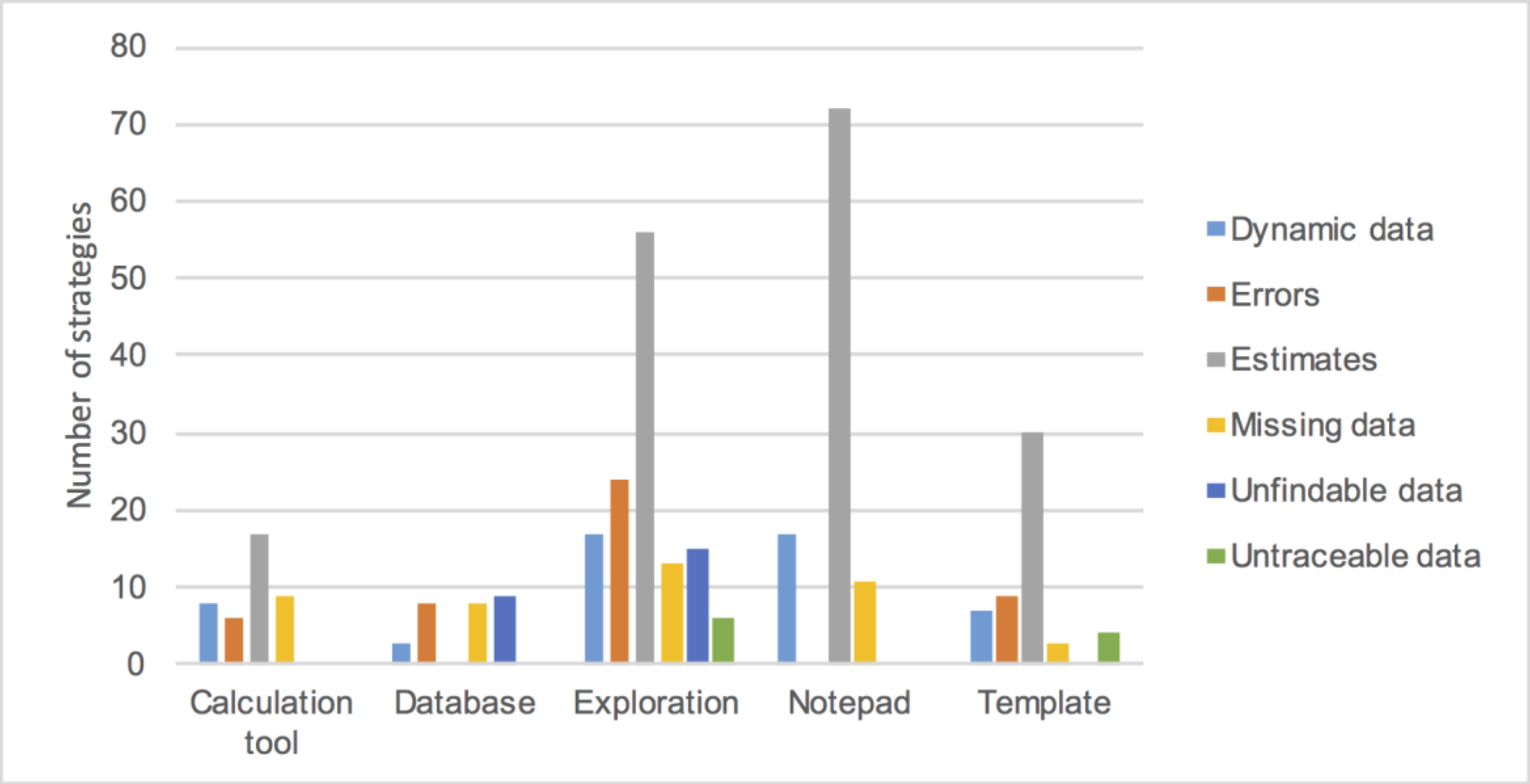

Figure 4. Bar chart showing the types of uncertainty per spreadsheet role.

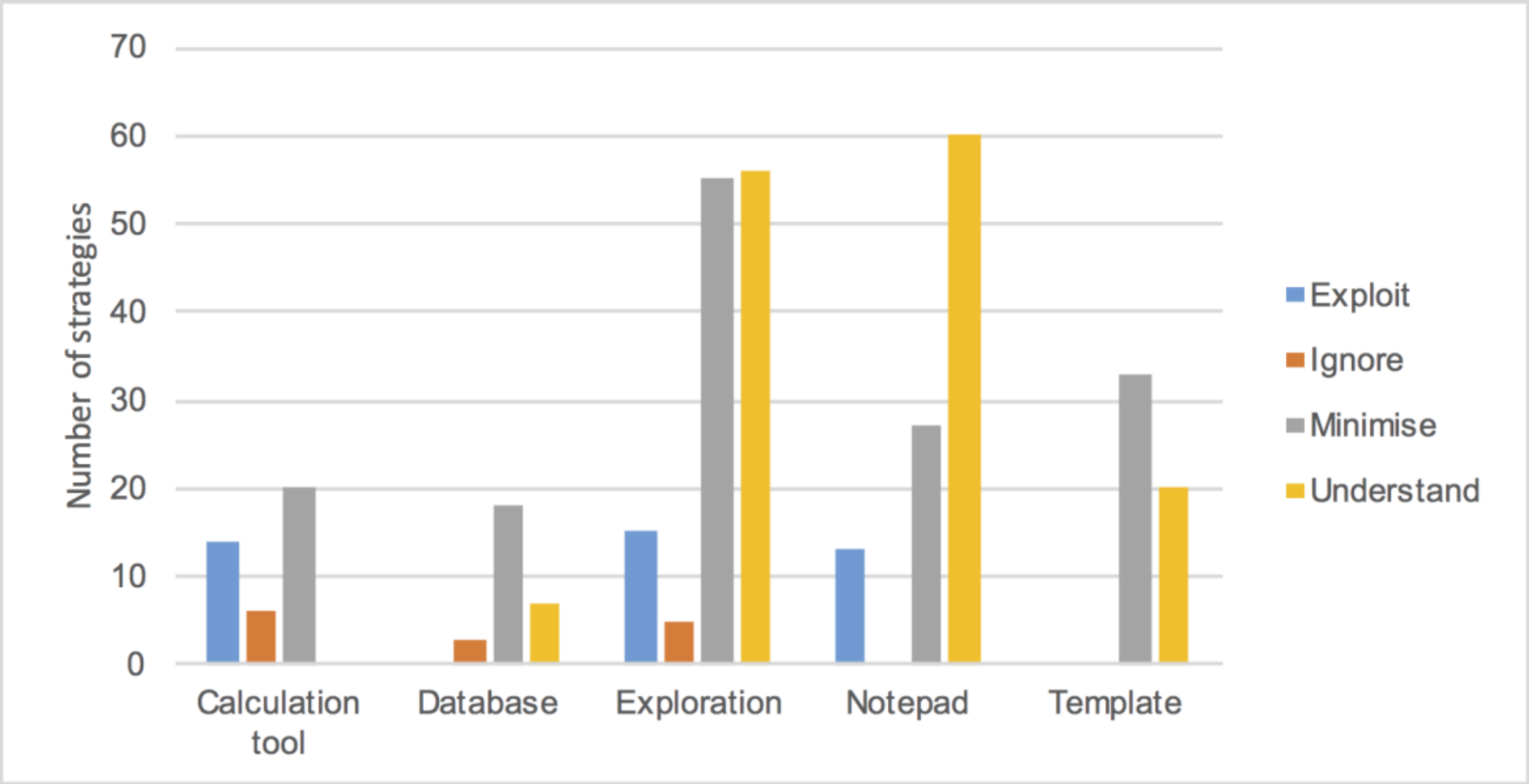

Figure 5. Bar chart showing the types of strategy goals per spreadsheet role.

| | Calculation tool | Database | Exploration |
|---|---|---|---|
| **Dynamic data** (10 participants) | **M:** *automatic updates (P6, P9);* update manually (P2, P3);<br>**E:** *Interact through model (P2, P5, P6, P9)* | **M:** *automatic updates (P6, P7, P9)* | **M:** replace with deterministic value (P3, P4, P7, P10, P11);<br>**U:** annotate level of uncertainty (P2, P5); plot (P6);<br>**E:** *consider multiple possible values (P2, P3, P4, P5, P6, P8, P9, P10, P11)* |
| **Errors** (9 participants) | **M:** replace with deterministic value (P10, P11); *rewrite dependencies spreadsheet (P6, P9);*<br>**I:** Remove (P10, P11) | **M:** acquire more data (P3, P5, P7);<br>**U:** discuss with colleagues (P10, P11); compare with own expectations (P1, P6, P9) | **M:** replace with deterministic value (P10, P11); acquire more data (P3, P7, P10, P11);<br>**I:** remove (P10, P11);<br>**U:** separate from rest of data (P10, P11); plot (P6, P9, P10, P11); annotate level of uncertainty (P2, P4, P10, P11); evaluate source of uncertainty (P10, P11); compare with own expectations (P3, P4, P10, P11) |
| **Estimates** (11 participants) | **M:** use domain specific prediction models (P2, P5, P6, P8, P9, P10, P11);<br>**E:** *Consider multiple possible values (P1, P2, P3, P4, P5, P6, P8, P9, P10, P11)* | - | **M:** compare with historical data (P2, P3, P4, P6, P8, P9); acquire more data (P1, P3, P5, P7, P9); replace with deterministic value (P3, P4, P6, P8); *present subset of scenarios (P1, P2, P3, P8, P9);*<br>**U:** evaluate source of uncertainty (P2, P3, P6, P8, P9); read and compare to literature (P2, P10, P11); compare with own expectations (P1, P2, P3, P4, P6, P9, P10, P11); compare with other forecasts (P2, P9); discuss with colleagues, experts (P3, P10, P11); *plot (P1, P2, P3, P5, P6, P8, P9, P10, P11;)*<br>**E:** *analyse subset of scenarios (P1, P2, P3, P8, P10, P11* |
| **Missing data** (7 participants) | **M:** replace with deterministic value (P6, P10, P11); update manually (P3, P7);<br>**I:** ignore (P10, P11); remove (P10, P11) | **M:** acquire more data (P2, P10, P11);<br>**I:** ignore (P6, P10, P11);<br>**U:** evaluate source of uncertainty (P10, P11) | **M:** replace with deterministic value (P6, P10, P11); acquire more data (P2, P10, P11);<br>**I:** remove (P10, P11);<br>**U:** separate from rest of data (P10, P11); evaluate source of uncertainty (P10, P11); plot (P6, P10, P11) |
| **Unfindable data** (7 participants) | - | **M:** *split up dataset into smaller sets (P3, P7, P9, P10, P11); write formulas to retrieve data (P1, P6, P7, P11)* | **M:** *split up dataset into smaller sets (P3, P9, P10, P11); write formulas to retrieve data (P6, P7, P10, P11); summarise data into smaller dataset (P7, P10, P11);* work with own estimate (P6, P11);<br>**U:** plot (P9, P11) |
| **Untraceable data** (7 participants) | - | - | **M:** acquire more data (P9, P11); work with own estimate (P1, P6, P11)<br>**I:** remove (P10) |

| | Notepad | Template |
|---|---|---|
| **Dynamic data** (10 participants) | **U:** explain uncertainty (P2, P5, P6, P8); **E:** *Consider multiple possible values (P1, P2, P3, P4, P5, P6, P7, P8, P9, P10, P11);* add variance (P3, P9) | **M:** *automatic updates (P6, P7, P9);* update manually (P2, P3, P5, P7) |
| **Errors** (9 participants) | - | **M:** *rewrite dependencies spreadsheet (P6, P7, P9);* acquire more data (P4, P7, P9) **U:** Evaluate source of uncertainty (P7, P10, P11) |
| **Estimates** (11 participants) | **M:** compare with historical data (P2, P3, P4, P6, P8, P9); use domain specific theories (P2, P6, P8, P9, P10, P11); adjust manually (P1, P3, P4, P9, P11); acquire more data (P2, P3, P4, P5); *give to colleagues to validate estimates (P3, P11);* **U:** report (P2, P3, P6, P11); justify estimates (P1, P2, P3, P8, P10, P11); present confidence (P1, P3, P4, P6, P10, P11); read and compare with literature (P2, P10, P11); compare with other forecasts (P2, P9); annotate level of uncertainty (P1, P2, P4, P5, P6); *comparison outcomes scenarios (P1, P2, P3, P4, P5, P6, P8, P9, P10, P11);* plot (P1, P2, P3, P5, P6); discuss with colleagues and experts (P3, P6, P11); explain uncertainty (P2, P3, P4, P6, P8) | **M:** compare with historical data (P2, P3, P4, P6, P8, P9); update manually (P3, P4, P7, P8); acquire more data (P2, P3, P5) **U:** evaluate source of uncertainty (P2, P4, P6, P8); read and compare to literature (P2, P10, P11); compare with other forecasts (P2, P9); compare with own expectations (P1, P2, P3, P4, P6, P9, P10, P11) |
| **Missing data** (7 participants) | **M:** update manually (P3, P7); replace with deterministic value (P4, P6); **U:** annotate level of uncertainty (P2, P4, P10, P11); report (P2, P10, P11) | **M:** acquire more data (P2, P3, P5) |
| **Unfindable data** (7 participants) | - | - |
| **Untraceable data** (7 participants) | - | **M:** *rewrite dependencies spreadsheet (P1, P6, P7);* acquire more data (P4) |

Table 2. Overview of uncertainty coping strategies, grouped by uncertainty type and the spreadsheet role. The rows represent the different types of uncertainty as described in section 4.1. The columns represent the different spreadsheet roles as explained in section 4.9. Labels M, I, U, and E indicate whether the strategy aimed to Minimise, Ignore, Understand or Exploit uncertainty. Strategies which are a new addition with respect to Boukhelifa et al’s framework are highlighted in italics.

## 5 DISCUSSION

The aim of this paper was to understand how people use spreadsheets to manage uncertainty. We confirm and extend previous findings on how people cope with uncertainty, and contribute novel insights into how the structure and form of spreadsheets are utilised to handle uncertainty: it is used as a database to hold historic and uncertain datasets, as a template to record and repeat data analysis, as a calculation tool to compute values and understand relationships between different variables, as a notepad to compare discrete scenarios, and as an exploration tool to get a better understanding of the uncertainty of a dataset or situation. Specifically, spreadsheets are used both to calculate uncertainty, but also to visually explore a situation's uncertainty, and as a data source when presenting uncertainty. Similar to prior work (Boukhelifa et al., 2017; Schunn and Trafton, 2012), we found various strategies to reduce uncertainty. However, we also found that, even participants who may not have a mathematical understanding of the uncertainty of their data, used visual properties of spreadsheets to explicitly represent uncertainty.

Prior work highlighted that one of the key features of spreadsheets is its malleability, and is used in situations where data has to be edited (Dourish, 2017). In our study, we found that when dealing with uncertainty, this malleability enabled participants to use spreadsheets as a notepad and exploration tool and explore and understand different possible scenarios. However, we also found that an important aim was to then transform the data and insights from this analysis stage into a form that can be presented to decision-makers, and spreadsheets provide limited support for this purpose. Spreadsheets were not only difficult to summarise as deliverables, but were also rarely shared in the analysis stage because of their comprehensiveness and complexity. Though spreadsheet content could be the result of collaborations outside of the spreadsheet, the structure was often the reflection of a single user's point of view, which was difficult to understand for others.

The spreadsheet grid and formulas enable users to structure data in a certain way and to perform calculations (Dourish, 2017). Indeed, we found that spreadsheets were used as a database to capture and structure data, and as a calculation tool and notepad to play around with multiple values. However, it was also used to get a visual understanding of the level of uncertainty. Because spreadsheets do not provide a visual distinction between certain and uncertain data, participants used various spreadsheet features to make this distinction, such as colours, formulas, comments, and spreading multiple possible values of uncertain data over the grid. These affordances both enable and constrain: it facilitates quick scanning of data to determine the level of uncertainty, and allows side-by-side comparison of scenarios. On the other hand, the time and space involved in creating scenarios motivated users to only consider a subset. Furthermore, by being constrained to input one value per cell, cells are handled by spreadsheets as separate discrete values, rather than relating to the same uncertain value.

### 5.1 Implications for design

Based on our findings, we discuss potential directions for imagining new spreadsheet functionalities to help users manage uncertain data:

- *Analysis and presentation layers.* Effort was involved in compressing comprehensive spreadsheets into summaries to use for decisions. Participants presented fixed numbers to clients and executives, but still retained the original uncertain data to refer to if they had to justify their presentation. Though the structure of spreadsheets was re-used, the content of spreadsheets was continually changing, which meant presentations had to be updated as well. It would therefore be valuable if spreadsheets would allow for easy switching between data analysis and presentation modes when handling uncertain data. Applications such as PowerPoint currently already support this switching between processes by providing an editing and presentation mode. A similar mode could be imagined for spreadsheets to make switching between processes within spreadsheets easier and straightforward.
- *Highlighting of uncertain data values.* We found that visual cues such as colour were used by participants to indicate uncertainty, as these were easier to cognitively process than more precise numbers. One can imagine spreadsheet functionalities to detect and highlight a cell that is likely to contain an uncertain value. For example, if a user edits a value multiple times, or has entered a formula rather than a discrete value, spreadsheets could recognise and highlight they are likely entering an uncertain value.
- *Direct input and manipulation of uncertain values.* Participants had workarounds to input uncertain values by spreading out multiple values over multiple cells. However, these were then seen and considered by spreadsheets as separate, discrete values, rather than multiple instances of the same variable. If spreadsheets would allow to input uncertain values directly into a cell, these could be used in further calculations, resulting in more accurate calculations and estimations. Furthermore, it should support different ways to input uncertain values. A number of participants considered uncertain values as distributions. Some had a range of possible values in mind, whereas others only knew a value was uncertain.

### 5.2 Limitations

Though we aimed to recruit a range of spreadsheet users with different skill levels, all participants who applied to take part were male, and handled uncertainty in spreadsheets for work purposes. This may have had an impact on their strategies: participants may have had different attitudes towards uncertainty, or learnt skills and workarounds through work to deal with uncertainty. However, uncertainty can play an important role in non-work related tasks as well. For future studies, it would be interesting to explore to what extent people use spreadsheets to manage uncertainty for personal tasks, such as managing a personal budget, or comparing mortgages.

## 6 CONCLUSION

This paper presented findings from an interview study on how spreadsheets are used to manage uncertainty. The findings identified five different spreadsheet roles: a database, template, calculation tool, notepad, and exploration tool. We found that the type of role impacts the strategies people adopt to manage uncertainty. Spreadsheets were used for calculation purposes to produce accurate estimates of uncertain values, but were also used to visually explore, understand and present a dataset or situation. Spreadsheets enabled but also constrained users in achieving their aims, and there was a trade-off between capturing data uncertainty and making data understandable and usable for others. This challenge introduces possibilities for new spreadsheet functionalities which could better support users to handle uncertainty.

## ACKNOWLEDGMENTS

We would like to thank our colleagues at Microsoft Research for many discussions and feedback, which were invaluable in shaping this research. A special thanks to Claudio Russo for his time and support. We would also like to thank Duncan Brumby for useful comments on an early draft of this work.